\documentclass[11pt]{article}
\usepackage{moriond,epsfig}
\usepackage{graphicx}
\usepackage{array}

\def\be{\begin{equation}}
\def\ee{\end{equation}}
\def\bea{\begin{eqnarray}}
\def\eea{\end{eqnarray}}

\def\piplus        {$\pi^{+}$}
\def\pimin         {$\pi^{-}$}

\def\kzeros        {K$^{0}_{s}$}

\def\lambdazero    {$\Lambda$}
\def\antilambda    {$\overline{\Lambda}$}
\def\ap            {$\overline{p}$}

\def\ximin         {$\Xi^{-}$}
\def\axi           {$\overline{\Xi}^{+}$}

\def\omegamin      {$\Omega^{-}$}
\def\aomega        {$\overline{\Omega}^{+}$}
\def\phikk         {$\phi(1020)$}

\def\munit{\ifmmode{\,\mathrm{MeV/{\mit c}^{\,2}}}
          \else{MeV/$c^{\,2}$}\fi}
\def\mup{\ifmmode{\mathrm{\,MeV/{\mit c}}}
          \else{MeV/{\it c}}\fi}
\def\mupp{\ifmmode{\mathrm{\,(MeV/{\mit c})^2}}
          \else{(MeV/{\it c})$^2$}\fi}

\def\gunit{\ifmmode{\,\mathrm{GeV/{\mit c}^{\,2}}}
          \else{GeV/$c^{\,2}$}\fi}
\def\pup{\ifmmode{\mathrm{\,GeV/{\mit c}}}
          \else{GeV/{\it c}}\fi}
\def\pupp{\ifmmode{\mathrm{\,(GeV/{\mit c})^2}}
          \else{(GeV/{\it c})$^2$}\fi}
\def\pum{\ifmmode{\mathrm{\,(GeV/{\mit c})^{-1}}}
          \else{(GeV/{\it c})$^{-1}$}\fi}
\def\pumm{\ifmmode{\mathrm{\,(GeV/{\mit c})^{-2}}}
          \else{(GeV/{\it c})$^{-2}$}\fi}

\def\ptt  {$p^2_t$}
\def\xf   {$x_{F}$}

\begin{document}
\vspace*{4cm}
\title{Strangeness Production at the HERA-B Experiment.}

\author{ M. Zavertyaev for the HERA-B collaboration }

\address{Max-Plank Institut f\"ur Kernphysik, Saupfercheckweg 1,\\
69117 Heidelberg, Germany.}

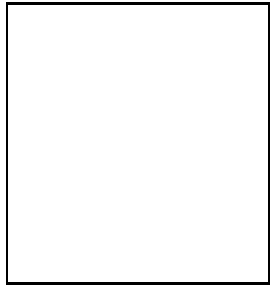
\maketitle\abstracts{
HERA-B is a fixed target experiment at the 920 GeV HERA proton beam 
at DESY which uses a variety of nuclear targets. During the last data
taking period from Nov. 2002 to Feb. 2003 , 200 million minimum bias 
events were recorded.
A large sample of $V^0$(\kzeros , \lambdazero, \antilambda),
approximately 20000 cascade hyperons $\Xi^{\mp}$ and 1200 $\Omega^{\mp}$
over low background were reconstructed using these data. 
About 1\,million $K^*(892) \rightarrow K\pi$ and 60,000 
$\phi(1020) \rightarrow K^+K^-$
decays in central production $(-0.15 < x_F < 0.1,\ 0.5 < p_t^2 < 12.1\,$\pupp$)$
were analyzed.
Preliminary results for differential \ptt\ spectra, $A$-dependence and
anti-particle to particle ratios are presented.}

\section{Introduction}

For more than 50 years, the production of strange particles was studied in a 
variety of beams for a wide energy range from few MeV up to 
$\sqrt{s_{NN}}\sim$ 200 GeV (see references in \cite{abt,hypv0,xiprod}).
The production of the strange quark and its subsequent 
hadronization in hadron-nucleus collisions constitutes an important benchmark 
test for QCD-based models such as PYTHIA, FRITIOF and
QCD-inspired phenomenological models \cite{NAZ89,NAZ92}
describing soft phenomena and for the applicability of perturbative QCD to hard
processes. But even now the comparison of experimental results with existing
models like \ PYTHIA and Quark Gluon String Model (QGSM) shows qualitative 
agreement only \cite{hypv0}.

A huge sample of minimum bias data collected in proton-nucleus collisions at
the HERA-B experiment permits to contribute to strange particle production
studies.
HERA-B is a fixed target experiment at the 920\,GeV proton storage ring HERA
at DESY.
It is a forward magnetic spectrometer with a large acceptance. 
It consists of a
high-resolution silicon vertex detector (VDS), drift chambers (DC),
a Cherenkov ring image detector (RICH), an electromagnetic calorimeter
(ECAL) and a muon detector \cite{abt}.  In this analysis, the information  
provided by the VDS, DC
and RICH detectors was used. The data were taken with carbon, titanium and
tungsten target wires. The present study uses a sample of about 170 million
events which was taken at mid-rapidity (\xf $\sim 0$). 

\section{Event selection}

Two different classes of events were selected for the analysis. The first class
is a data sample with long living particles like $V^0$ and the cascade hyperons
$\Xi^{\mp}$ and $\Omega^{\mp}$. The second class includes the short living 
resonances $K^*(892)$ and \phikk\ selected by using particle identification 
by the RICH. 

Only events with exactly one reconstructed primary vertex (of at least two 
tracks) were accepted for the analysis.
In each event with at least four tracks, a full combinatorial search for
$V^0$ candidates was performed. All pairs of oppositely charged tracks with
a minimum distance smaller than \mbox{0.07\,cm} are considered. 
Clear signals of the decays of
\kzeros , \lambdazero\ and \antilambda\ are seen above a high background.
The  additional cut of $p_t\cdot c\tau > 0.05$\pup$\cdot cm$
removes about 90 \% of the background while the signal losses are below
5 \%. No further cut on particle identification is applied. The invariant mass
distributions of the \piplus \pimin ,  $p$\pimin\ and \ap\,\piplus\ 
combinations are shown in Fig.\,1.

The \lambdazero (\antilambda) candidates are accepted for further analysis if 
the invariant mass belongs to the region of $\pm  3 \sigma$ around the peak
position ($ 1.11  \gunit  < m < 1.121 \gunit $ ). The contamination from 
\kzeros\  decays is suppressed by a cut of $\pm  3 \sigma$ around the \kzeros\ 
mass in the invariant mass spectrum 
($ 0.482 \gunit <m_{\pi^- \pi^+ }<0.512 \gunit $).
\begin{figure}[t] 
\begin{minipage}[c]{2.5in}
\includegraphics[width=2.5in]{fig.1a}
\caption{The invariant mass distributions for \piplus \pimin, $p$\pimin\
         and \ap\,\piplus. $\Delta$ m = 1.0/0.4\munit\ in a)/b),c).}
\label{fig:v0mass}
\end{minipage}
\hspace{0.4cm}
\begin{minipage}[c]{3.2in}
\vspace{-0.4cm}
\includegraphics[width=3.5in]{fig.1b}
\caption{The invariant mass distributions for cascade hyperons 
($\Xi$, $\Omega$) and resonances $K^*(892)$ , \phikk\ . }
\label{fig:cass}
\end{minipage}
\end{figure} 


The invariant mass of the \lambdazero (\antilambda) candidates combined with 
one additional negative (positive) track is shown in 
Fig.\,\ref{fig:cass}a,b when this track is considered as a pion or a kaon.
Clear signals of cascade hyperons are obtained by requesting that none of
the decay products but only the  $\Xi^{\mp}$ or $\Omega^{\mp}$ itself must 
point to the primary vertex.

Signals of one million of $K^*(892)$ decays  \cite{christofer} and about 
$5\cdot 10^4$ \phikk\ decays \cite{mitch}
were observed (Fig.\,\ref{fig:cass}c,d) in the invariant mass
distribution of oppositely charged tracks emerging from the primary vertex
with the cut on the kaon identification likelihood
given by the RICH \cite{rich}. The cut on likelihood values were set at 0.3 
for both kaons coming from the \phikk\ decay and 0.95 in case of $K^*(892)$. 
The low momentum limit for both kaons coming from the \phikk\ decay was 
set at 10\pup .

\section{Results}

The analysis of the data covered the
differential \ptt - spectra, dependency on the atomic mass of the target 
nucleus and anti-particle to particle ratios.
\vspace{0.5cm}

\parbox[t]{4.cm}
{
{Table 1: The parameter B of the $p^2_t$ spectra.}
\setlength{\extrarowheight}{3pt}
\begin{tabular}{|c|c|}
\hline
& B, \pumm\\ \hline
\kzeros     & 3.6\ $\pm$0.1 \\
\lambdazero & 2.4\ $\pm$0.1 \\
\antilambda & 2.2\ $\pm$0.1 \\
\ximin      & 1.8\ $\pm$0.2 \\
\axi        & 1.6\ $\pm$0.2 \\
\omegamin   & 0.95$\pm$0.4 \\
\aomega     & 0.93$\pm$0.4 \\
\hline
\end{tabular}
}
\hspace{1.cm}\parbox[t]{10.3cm}
{
For \kzeros\ the observed \ptt\,spectrum is shown in
Fig.\,\ref{fig:pt}. For the other particles, the \ptt\,spectra look very 
similar. All distributions may be described by the simple Gaussian
formula $d\sigma / dp^2_t \propto exp(-B \cdot p^2_t)$
up to 0.8-1\pupp\ approximately. At higher \ptt\ values the spectra become 
flatter. The results of the fits are shown in Table 1. No significant 
difference was observed between the B values obtained fron the different 
targets. The measured 
B values follow the general trend observed by other experiments \cite{hypv0} - 
the higher the mass of the particle, the wider the \ptt\,spectrum.

The \ptt\,spectra for \phikk\ and  $K^*(892)$ were measured up to 12\pupp . A 
good approximation for the whole \ptt\ range of these spectra is 
$d\sigma / dp^2_t \propto (1+ p^2_t / p^2_0)^{-\beta} $
in contrast to the Gaussian ansatz at low \ptt . The value of $\beta $
for $K^*(892)$ decreases from 6 to 4.5 for C:Ti:W targets while for \phikk\
it stays constant $\approx$4.3  for all targets. The errors on the $\beta $ 
values are from 5 to 8\% .
}
\vspace*{-0.5cm}
\begin{figure}[htb] 
\begin{minipage}[c]{2.3in}
\includegraphics[width=2.5in]{fig.2a}
\caption{The differential \ptt\ production spectrum for \kzeros\
         on carbon nuclei.}
\label{fig:pt}
\end{minipage}
\hspace{0.5cm}
\begin{minipage}[c]{3.5in}
\vspace{0.3cm}
\includegraphics[width=3.2in]{fig.2b}
\caption{The ``Cronin'' effect for \phikk\ and $K^*$.}
\label{fig:cronin}
\end{minipage}
\end{figure} 


Fitting the \ptt\,spectra of the three materials for \phikk\ and  $K^*(892)$ 
in each  momentum intervall with the conventional formula 
$d\sigma_{pA} / dp^2_t =$ $A^{\alpha} \cdot d\sigma_{pp} / dp^2_t$ 
yields the exponent $\alpha$ to rise above unity at high \ptt\ 
(Fig.\,\ref{fig:cronin}). This is the so-called 
Cronin effect which is well known for stable baryons \cite{cronin}.
Now effect is clearly observed also for the \phikk\ and $K^*(892)$ resonances.

\begin{figure}[htb] 
\centering
\includegraphics[width=15.cm]{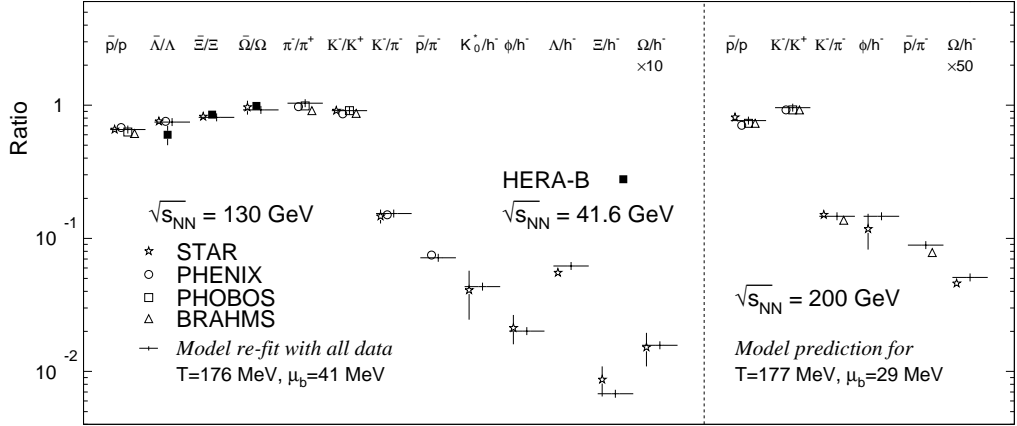}
\vspace{-1.cm}
\caption{The anti-particle to particle ratio measured at HERA-B in 
          comparison with the pulished data \protect\cite{ion}.}
\label{fig:ratio}
\end{figure}

\begin{figure}[htb] 
\addtolength{\abovecaptionskip}{-50pt}
\centering
\includegraphics[width=7.cm]{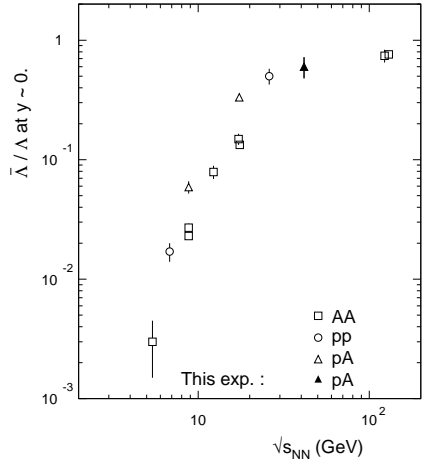}
\caption{The anti-particle to particle ratio versus energy 
  \protect\cite{abt}.}
\label{fig:lratio}
\end{figure}

The anti-particle to particle ratios at HERA-B were measured for $V^0$s and
cascade hyperons. The results are shown in Fig.\,\ref{fig:ratio}. In the same 
plot results in Au-Au-collisions from different heavy-ion experiments are
shown for comparison. The \antilambda /\lambdazero\ ratio at HERA-B energy 
is about 20\% smaller than at RHIC. In Fig.\,\ref{fig:lratio}, this ratio is 
shown as a function of the energy. At HERA-B energies
this is a rising function and at least a part of the 20\% difference may be 
attributed to the energy dependence and not exclusively to the difference
between $pA\ -\ AA$ interactions.

\section{Conclusion}

The minimum bias data sample taken by the HERA-B spectrometer
is a source of high statistic signals of strange particles decays.
On its basis studies of the \ptt\ differential production spectra were
performed. The A-dependence studies show the ``Cronin'' 
effect for the \phikk\ and $K^*(892)$ resonances. The comparison of the 
anti-particle to particle ratios in $pA$ with the ratios in $AA$ interactions
does not show a strong difference. 

\vspace{-0.8cm}

\vspace*{-1.cm}


\end{document}